\documentclass[12pt]{iopart}

\usepackage{iopams}  
\usepackage{float}
\usepackage{graphicx}
\usepackage{amssymb}
\usepackage{hyperref}
\usepackage[utf8]{inputenc}
\usepackage{upgreek}

\begin{document}

\title[]{Harnessing of temporal dispersion for integrated pump filtering in spontaneous heralded single-photon generation processes}

\author{Julian Brockmeier$^1$, Timon Schapeler$^{1,2}$, Nina Amelie Lange$^{1,2}$, Jan Philipp Höpker$^1$, Harald Herrmann$^3$, Christine Silberhorn$^3$ and Tim J. Bartley$^{1,2}$}

\address{$^1$Department of Physics, Paderborn University, Warburger Str. 100, 33098 Paderborn, Germany
$^2$Institute for Photonic Quantum Systems (PhoQS), Paderborn University, Warburger Str. 100, 33098 Paderborn, Germany
$^3$Integrated Quantum Optics Group, Institute for Photonic Quantum Systems (PhoQS), Paderborn University, Warburger Str. 100, 33098 Paderborn, Germany}
\ead{julian.brockmeier@upb.de}
\vspace{10pt}

\begin{abstract}
Cointegration of heralded single-photon generation and on-chip detection requires the ability to differentiate between pump light and single photons. We explored the dispersion-induced temporal separation of optical pulses to reach this goal. Our method exploits the distinct group velocities of pump light and single photons, as well as single-photon detectors with high timing resolution. We simulate the propagation for photon pair generation by spontaneous parametric down-conversion in titanium in-diffused waveguides in lithium niobate and thin-film lithium niobate, and spontaneous four-wave mixing in silicon on insulator and silicon nitride. For the different integration platforms, we show the propagation distance required to sufficiently distinguish between pump and single photons for different timing resolutions, and demonstrate that this should be feasible with current superconducting nanowire single-photon detector technologies.
Finally, we experimentally simulate our approach using the dispersion in the optical fiber.
\end{abstract}


\submitto{\NJP}
\maketitle

\section{Introduction}
Quantum photonics has made significant contributions in recent years in fields such as quantum computation~\cite{Knill2001,Slussarenko2019,Zhong2020,Madsen2022}, quantum communication~\cite{Gisin2007}, and quantum metrology~\cite{Giovannetti2011,Polino2020,You2021}. The integration of multiple photonic elements provides many advantages, such as stronger nonlinear interactions, smaller footprint, lower interface losses, and greater circuit complexity~\cite{Elshaari2020, Moody2022, Wang2020}. The key elements of integrated quantum photonics are the generation, manipulation, and detection of single photons. An important class of single-photon sources is based on heralded spontaneous nonlinear optical processes, such as spontaneous parametric down-conversion (SPDC) and spontaneous four-wave mixing (SFWM)~\cite{Eisaman2011}. Both processes have been shown to be cryogenically compatible~\cite{lange2022cryogenic, lange2023degenerate, feng2023entanglement, cheng2024efficient}, which allows for integration with superconducting detectors that require cryogenic operation~\cite{you2020superconducting, Elshaari2020}. These two methods to generate single photons require pump light which can spontaneously decay into photon pairs, whereby measuring one photon heralds the existence of the other. These processes have a low generation probability in the order of~10\textsuperscript{-10}~\cite{Hoz2020}, {i.e.} only a small fraction of the pump light spontaneously converts. In order to measure the down converted photons, the pump must be suppressed by more than 100 dB~\cite{Montaut2017, Gentry2018}.

In general, single-photon-level detectors are typically not able to resolve the wavelength of the incident light. Therefore, in order to differentiate the pump light from the single-photon pairs, a wavelength-selective component is needed. For free-space applications, this is easily achieved by filters; however, in integrated quantum photonics, this is highly challenging. Scattering and/or coupling to substrate modes are limiting factors when fabricating integrated wavelength-selective devices, which become significant noise sources because of the close physical proximity of sources and detectors. 
 Table \ref{tab:SuppressionOverview} shows the performances of various approaches to integrated wavelength-dependent pump suppression in different waveguide platforms, namely silicon on insulator (SoI), silicon nitride (SiN), titanium-in-diffused waveguides in lithium niobate (Ti:LN) and thin-film lithium niobate (TFLN). 

\begin{table}
\caption{\label{tab:SuppressionOverview}Overview of various pump rejection methods and their yield for SiN, SoI, TFLN and Ti:LN.}
\begin{tabular}{lll}
\br
Platform & Technology                                      & Pump rejection                                               \\
\mr
SoI      & directional couplers                           & 17 dB ~\cite{Magden2018}                        \\
SoI      & balanced Mach Zehnder interferometer           & 15 dB ~\cite{Horst2013}                      \\
SoI      & cascaded Mach Zehnder interferometer           & \textgreater 100 dB ~\cite{Pikarek2017}      \\
TFLN     & ring resonator                                 & 25 dB ~\cite{Bahadori2019}             \\
SoI      & cascaded micro ring resonators                 & \textgreater 95 dB~\cite{Gentry2018}, \textgreater 100 dB~\cite{Cabanillas2021}  \\
SoI      & Bragg reflector                                & 65 dB~\cite{Harris2014}                \\
SiN      & cascaded grating assisted directional couplers & 68.5 dB~\cite{Nie2019}                 \\
SiN      & ring resonator                                 & \textgreater 95 dB~\cite{Elshaari2017} \\
Ti:LN       & directional couplers                           & 20-30 dB~\cite{Jin2014,Solntsev2017}\\ 
\br
\end{tabular}
\end{table}
 






All these approaches rely on spatial separation of different wavelengths, which may be limited by fabrication tolerances and can cause scattering. By contrast, relying on the intrinsic material properties may circumvent these issues. In this paper we investigate the possibility of filtering the pump light from the single photons by wavelength-selective dispersion. Using this method, the signal and pump light are separated in time due to their different propagation speeds. Using a detector with sufficient timing resolution, the pump and single photon can be distinguished by their arrival time. 

This paper is structured as follows. First, we describe the concept of dispersive pump rejection. Then, we simulate the relative dispersion of pump light and single photons in various integrated platforms, namely Ti:LN, TFLN, SoI, and SiN. We then experimentally simulate the concept of this filter by using a fiber loop representing an integrated resonator and an SPDC source in order to show a separation of the propagating pulses over time. 

\section{Dispersive pump filtering}
\begin{figure}[t]
\centering\includegraphics[width=13.2cm]{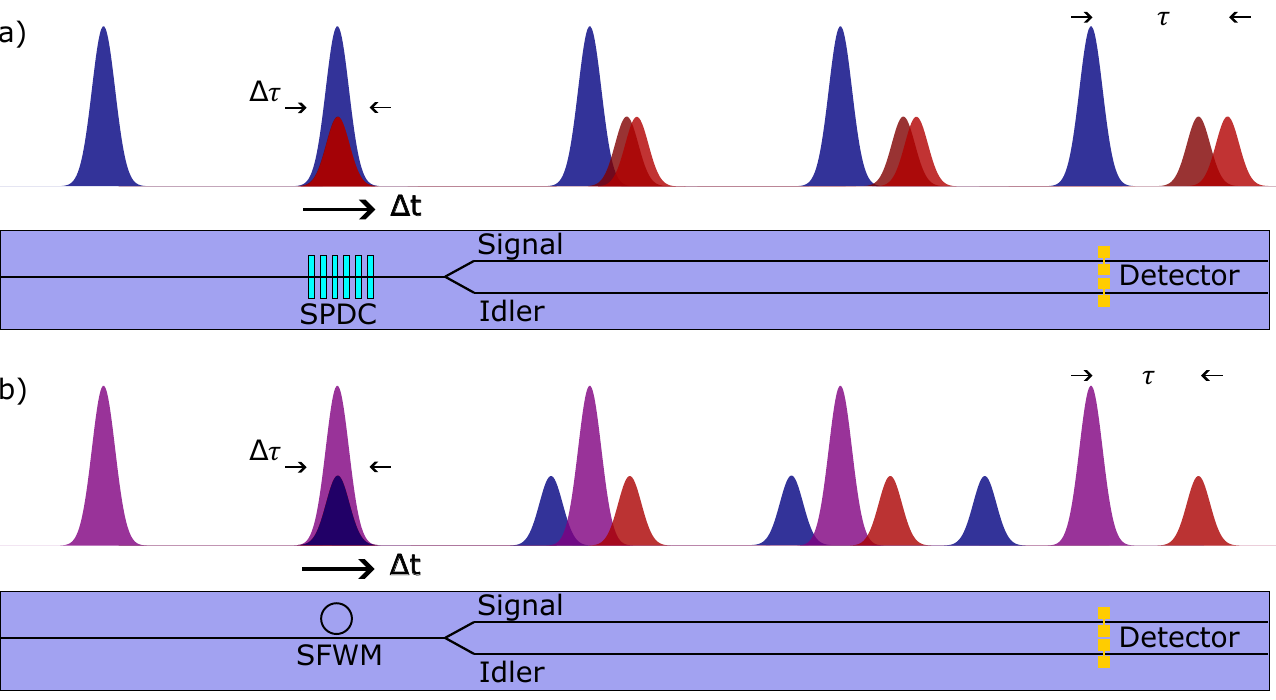}
\caption{Visualization of time based filtering for (a) spontaneous parametric down-conversion (SPDC) and (b) spontaneous four-wave mixing (SFWM). At the start of the SPDC process the single photons (red) are completely enveloped by the pump pulse (blue). 
Over time (from left to right) the pulses separate due to the different effective refractive group indices of the wavelengths and therefore their different propagation speeds. For the SFWM process the signal photon (red) is ahead of the pump light (purple), while the idler photon (blue) arrives last at the detector.}
\label{fig:idea}
\end{figure}

Dispersive pump filtering relies on chromatic dispersion due to different effective group indices for each wavelength. We assume normally dispersive media, which means that lower energy photons travel faster, and arrive at the detector earlier, compared to photons with higher energy. For SPDC~\cite{Eisaman2011}, this means that both signal and idler photons arrive earlier than the pump pulse, whereas for SFWM~\cite{Eisaman2011}, the longer wavelength photon must be used as the herald, since it arrives before both the pump and the shorter wavelength photon. Thus, we achieve a resolvable separation of the pump-photon pulses from at least one of the spontaneously generated single photons in the time domain, as depicted schematically in figure~\ref{fig:idea}. 

The filtering relies on the ability of the detector to resolve the arrival time of the photons relative to each other. This means the timing resolution $\Delta t$ (and optical pulse duration $\Delta\tau$) must be shorter than the difference in arrival time of the pump and single-photon pulse $\tau$, while the dead time of the detector must be shorter than the pump pulse separation (inverse of the repetition rate of the experiment). For ultrafast heralded photon sources, optical pulse durations of $\Delta\tau\lesssim$~1~ps are typical, whereas the system jitter of superconducting nanowire single-photon detectors (SNSPDs), including a clock signal, is typically $<$~50~ps, thus dominating the overall resolution of this approach. 


This method relies on sufficiently long propagation distances in order to maximize the separation of the pulses traveling at different velocities, as well as a large difference in the effective refractive group indices in the material. 
Therefore, it is also important that the medium has very low propagation losses. 


For evaluating the feasibility of dispersive pump filtering, we analyze the distinguishability between a detection event arising from the heralding photon and events from the pump photons. Our key figure of merit is the detection probabilities arising from detecting pump and single photons: we seek the arrival times at which the detection probability from single photon events dominates that from the pump photons. To do so, we investigate the fraction of the single-photon probability which overlaps with the pump probability, assuming a process with 10\textsuperscript{10} pump photons for each photon pair. A signal to noise ratio of 1 corresponds to the case where, given a detection event, the probability that this arose from a single photon or the pump are equal. This is equivalent to 100~dB spectral suppression of the pump, since the pump is 10 orders of magnitude higher in intensity. We take a stricter definition of separation to be the point at which the pump probability falls below 1~\% of the detection events, which corresponds to an equivalent spectral pump suppression of over 120~dB. 

\begin{figure}[t]
\centering\includegraphics[width=13.2cm]{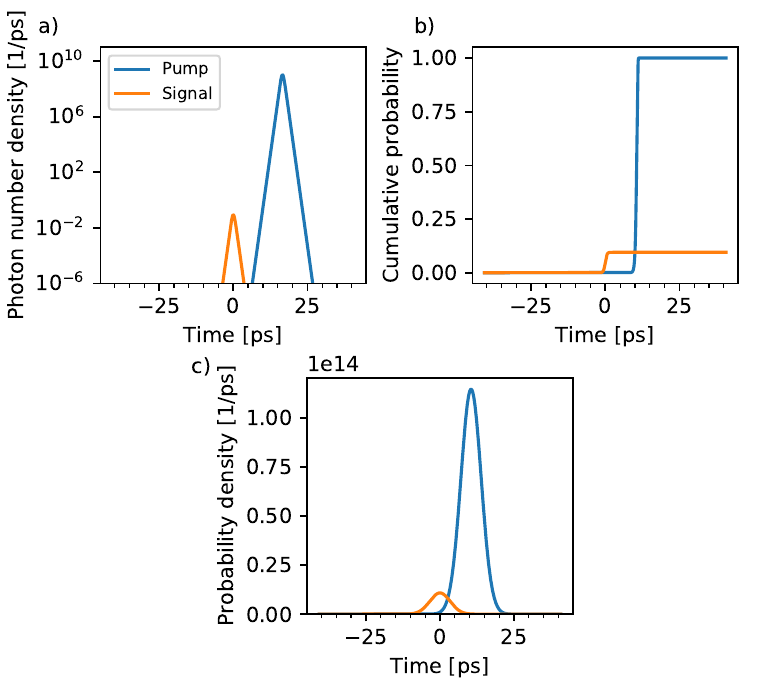}
\caption{Visualization of single-photon signal (orange) and pump signal (blue) in Ti:LN after the pulses propagated through the waveguide for 200~ps. Here (a) represents the photon number density of the 1 ps wide pulses, (b) shows the cumulative probability of detecting a click with a 100~\% detection efficiency and (c) shows the probability density while taking the detection jitter of 8~ps into account.}
\label{fig:PropOverview}
\end{figure}

To calculate this overlap, we start by defining the photon number density of the pulses as sech\textsuperscript{2}-shaped pulses
\begin{equation}
    \label{eq:gaussian}
    S(t)= \frac{A}{2 \sigma} \mathrm{sech}^{2} \left(\frac{t-t_\mathrm{c}}{\sigma}\right)\,,
\end{equation}
where the pulse area $A=\mu$, the mean photon number, and the full width at half maximum (FWHM). 
We define the FWHM as the pulse width of the photon signal, which is shown in figure~\ref{fig:PropOverview}~(a). Therefore the standard deviation of the function is $\sigma=\mathrm{FWHM}~2\sqrt{2\log(2)}$.

Here we assume 1~ps pulse duration. The center of the peaks $t_\mathrm{c}$ in the time domain are given by the relative propagation distance between the single photon and pump signal. This distance depends on the propagation length of the pulses as well as the corresponding effective refractive group index of the two pulses. 

Since usually only single-photon pairs are of interest, we set the number of pump photons to achieve a photon-pair generation probability of 10~\% per pump pulse in order to reduce the possibility of multi-photon components. The required number of pump photons per pulse to achieve this generation rate varies depending on the nonlinear medium. This number can be calculated from the inverse of the probability of a pump photon to decay into a photon pair via SPDC, of the order 10\textsuperscript{10} photons. Since four-wave mixing requires a similar amount of pump photons~\cite{Hoz2020}, we take this as a lower limit. 


The signal (and idler) pulse thus comprise 0.1 photons on average, from which their photon number density is calculated. As all pulses propagate, their amplitude reduces due to wavelength- and material-dependent losses, which we account for when calculating the overlap.

Assuming unit detection efficiency, the probability $P$ of a detector click is calculated using the probability mass function for the Poisson distribution
\begin{equation}
    \label{eq:poisson}
    P(\mu,k) = \frac{\mu^k e^{-\mu}}{k!}\,,
\end{equation}
where $k$ is the specific number of photons and $\mu$ is given as the mean photon number~\cite{migdall2013}. Assuming unit detection probability when at least one photon is incident on the detector, the probability that any photon is detected is given by $P_{\mathrm{det}}=1-P(\mu,0)$.

To calculate $\mu$, we use the temporal photon density given by the sech\textsuperscript{2}-shaped distribution of the signal pulse relative to its arrival at the detector as seen in equation~\ref{eq:gaussian}, such that integrating this distribution with respect to time gives the mean photon number $\mu$. By choosing the limits of the integration appropriately, the number of photons in our system observed by the detector becomes time-dependent, i.e., $\mu \rightarrow \mu(t)$. Consequently, we are able to determine the cumulative probability of detecting a photon after a certain time $t$. This results in a time-dependent detection probability  $P_{\mathrm{det}} \rightarrow P_{\mathrm{det}}(t)$. An increasing time-interval $\Delta t$ results into more photons arriving at the detector and therefore an increase of the detection probability, which is shown in figure \ref{fig:PropOverview}~(b). The cumulative probability saturates at $1-P(\mu,0)$ for the pump, which is essentially unity, and 0.1 for the signal, the probability of a signal photon generated in that pulse.

Since the signal and pump photons propagate at different speeds, the relative probability of measuring a signal photon compared to a pump photon changes as a function of propagation time. In order to calculate the point where the probability of the detection of single photons is distinct from the pump photons, we take the derivative of the cumulative probability to give the probability density. However, this probability is broadened by the jitter of the instrumentation (detector and clock), which adds an uncertainty of the detection events and therefore affects the distinguishability of the photons. We consider this by taking the convolution of the probability density, {i.e.}, the derivative of the cumulative probability distribution, and a Gaussian distribution $G_{\mathrm{jitter}}(t) = \frac{A}{\sigma \sqrt{2 \pi}} e^{-\frac{1}{2} (\frac{t-t_\mathrm{c}}{\sigma})^2}$ with the FWHM equal to the instrumentation jitter which is based on the standard deviation $\sigma$ and normalized to unity
\begin{equation}
    \label{eq:convolution}
    \left(\frac{\partial P_{\mathrm{det}}(t)}{\partial t} \ast G_{\mathrm{jitter}}\right)(t)  = \int\limits_{-\infty}^\infty \frac{\partial P_{\mathrm{det}}(\tau)}{\partial \tau}G_{\mathrm{jitter}}(t-\tau)\,d\tau\,,
\end{equation}
which is shown in figure \ref{fig:PropOverview}~(c). We consider single-photon detectors such as SNSPDs, which have demonstrated a jitter in the few picosecond regime~\cite{Korzh2020}. Recent integrated SNSPDs have shown a timing resolution $<$~20~ps~\cite{Lomonte2021, Beutel2022}. 


Integrating the probability density (see figure \ref{fig:PropOverview}~(c)) over a specific time window corresponds to the detection probability in that time. We take a 3$\mathrm{\sigma}$ integration window, which is equivalent to measuring 99~\% of the intensity of the single-photon signal.

Within this window, we are interested in the relative probability of measuring a click arising from the single photon, compared to clicks arising from the pump pulse. Reducing this overlap to 1~\% corresponds to 120~dB extinction. The point at which this extinction is reached will depend on the propagation distance and instrumentation jitter.


\subsection{Simulating dispersive pump filtering for integrated platforms}
We investigate the feasibility of dispersive pump filtering in four waveguide platforms: titanium in-diffused lithium niobate (Ti:LN), thin-film lithium niobate (TFLN), silicon on insulator (SoI) and silicon nitride (SiN). In each case, various wavelength combinations for the pump and generated single photons can be used, for which we assume a pulse-length of 1~ps. For Ti:LN, we consider Type-II degenerate SPDC pumped at 775~nm, which generates photon pairs in orthogonal polarization modes at 1550~nm~\cite{Montaut2017}. For TFLN we choose a Type-0 SPDC process at the same wavelengths where the pump light as well as the signal and idler photons are extraordinary polarized~\cite{Yusof2019}. 
For SFWM, we consider SiN pumped at 1540~nm, with single photons at 1600~nm \cite{Vitali2024}. For SoI we choose a process which is pumped at 1550 nm, generating photons at 1202~nm and 2181~nm~\cite{Signorini2018}. 
In general the effective refractive group index depends strongly on the waveguide parameters, which vary significantly between each platform and waveguide geometry. In the case of Ti:LN and TFLN, effective group refractive indices can be extracted from simulations~\cite{Bartnick2021,Babel2024}. However, relevant material parameters for SoI and SiN from which the effective group index can be derived depend strongly on waveguide geometry and are not routinely published. Nevertheless, for the sake of argument and proof of concept, we calculate the group index based on bulk material parameters in SoI and SiN; our results would need to be adapted for specific geometries for a stricter quantitative comparison. The material parameters we use are summarized in table~\ref{tab:ParameterOverview}.

\begin{table}
\caption{\label{tab:ParameterOverview}Overview of spontaneous single-photon processes and their corresponding wavelengths $\lambda_\mathrm{S}$ as well as the pump photon wavelength $\lambda_\mathrm{P}$ and their polarization (pol) and effective refractive group indices n\textsubscript{eff}.}
\begin{tabular}{lllllll}
\br
Platform & Process & $\lambda_\mathrm{P}$ (pol) [nm] & $\lambda_\mathrm{S}$ (pol) [nm] & n\textsubscript{g,P} & n\textsubscript{g,S}   & loss [dB/cm]                           \\
\mr                           
SoI      & SFWM    &  1550 (TE)                             & 1202 (TE) \cite{Signorini2018}            & 1.4626                & 1.4617 \cite{Malitson1965}  & 0.1  \cite{Chandra2019}   \\
SiN       & SFWM    & 1540 (TE)                           & 1600 (TE) \cite{Vitali2024}         & 2.0396               & 2.0395 \cite{Luke2015}  & 0.00045 \cite{Bauters2011}  \\
Ti:LN       & SPDC    & 775 (TE)                         & 1550 (TM)                        & 2.369                 & 2.187 \cite{Bartnick2021}                   & \textless 0.03  \cite{Hoepker2021} \\
TFLN     & SPDC    & 775 (TE)                         & 1550 (TE) \cite{Yusof2019}                         & 2.331                & 2.270  & 0.27   \cite{Luke2020}\\
\br
\end{tabular}
\label{tab:parameters}
\end{table}

\begin{figure}[t]
\centering\includegraphics[width=13cm]{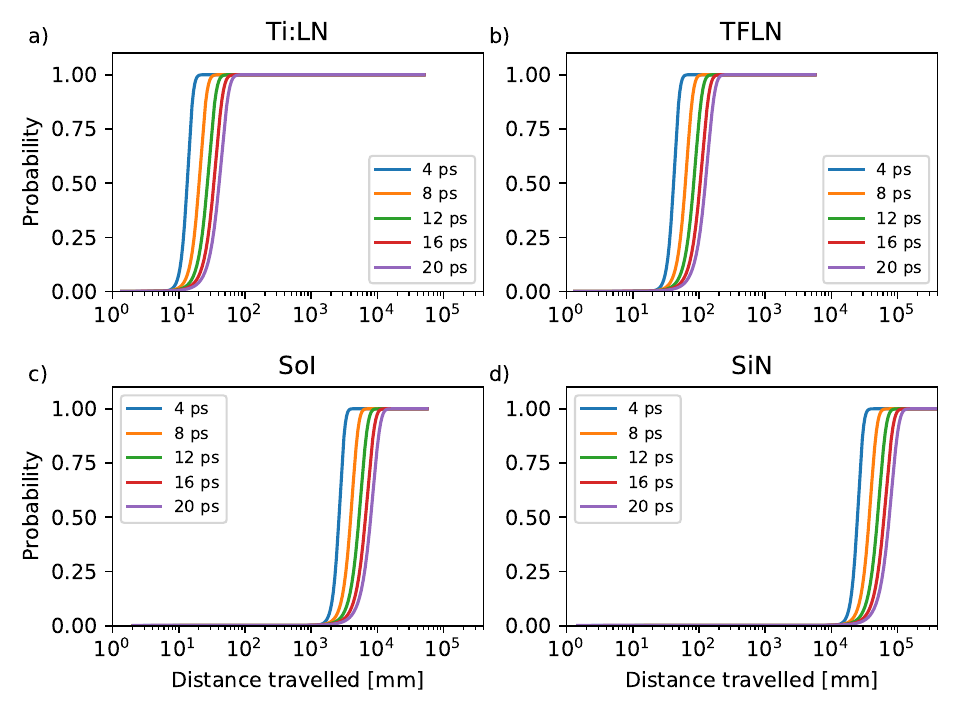}
\caption{Given a detection event, we plot the probability that the event arose from the signal photon, for the platforms (a) Ti:LN, (b) TFLN, (c) SoI, and (d) SiN, 
plotted as a function of propagation distances, for five different values of instrumentation jitter from 4~ps to 20~ps.}
\label{fig:Simulation}
\end{figure}

By comparing the probability $P_\mathrm{filteredSignal}=\frac{P_\mathrm{signal}}{P_\mathrm{cumul}}$ that the detected photon belongs to the signal photons and not to the pump pulse, we can simulate the point where both pulses are separated and thus become distinguishable, see figure~\ref{fig:Simulation}. This is done by taking the probabilities of click events based from pump-light into account thus resulting in the probability of $1-\frac{P_\mathrm{pump}}{P_\mathrm{cumul}}$. High propagation losses (see table \ref{tab:parameters}) have a strong impact for the single-photon signal, while there are still a lot of pump photons remaining in the system. Therefore, the losses are most prominent for SoI caused by the long travel distance.

\begin{figure}[t]
\centering\includegraphics[width=13cm]{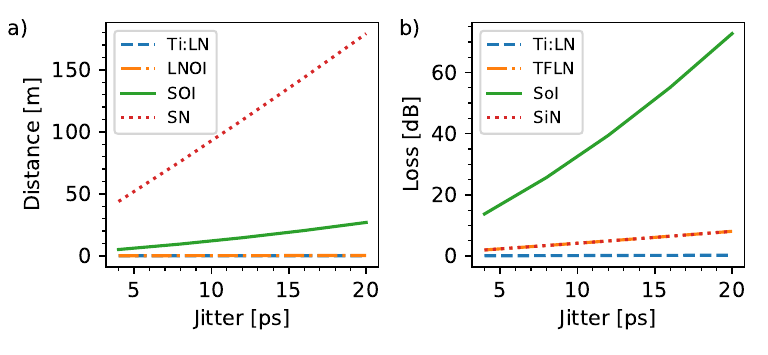}
\caption{ The distances (a) and corresponding losses (b) for the platforms  Ti:LN, TFLN, SoI, and SiN after the pulses reach a separation with an overlap of the detection probability below 1~\% for a different jitter of the SNSPD from 4~ps up to 20~ps.} 
\label{fig:Separation}
\end{figure}

For an instrumentation jitter of 20~ps, for TFLN we see a separation of the pulses after 299.30~mm which corresponds to propagation losses of 8.08~dB, which is shown in figure~\ref{fig:Separation}. For this platform integrated delay loops for single wavelengths have already been realized~\cite{Ekici2023}, which highlights the feasibility of dispersion based pump filtering. Another platform for SPDC processes is Ti:LN with a separation of the TM polarized single photon from the pump light after 90.08~mm. This is equivalent to propagation losses of 0.27~dB. In Ti:LN, the weak confinement limits the bend radius and thus fabrication of meandering structures. Therefore, the propagation distance is limited by the size of the chip.  Overall, the simulations show promising results for a dispersion based filter in scenarios where the single photons outpace the pump light. In particular, the development of integrated detectors with low jitter could further improve the feasibility of this approach. 

For SoI, the pulses become separated after 26.928~m, which corresponds to propagation losses of 72.71~dB. For SiN we see a separation of pump light and single photons after 179.2~m, as seen in figure~\ref{fig:Separation}. The corresponding propagation losses for the signal are 8.06~dB~\cite{Bauters2011}. The propagation distance may be possible with optical buffers \cite{Liu2019}. 
The simulations show that a physical separation of single photons from the pump light by dispersion alone is feasible for integrated photonic platforms.




\section{Experimentally simulating dispersive pump filtering using a fiber loop}

\begin{figure}[t]
\centering\includegraphics[width=1.0\textwidth]{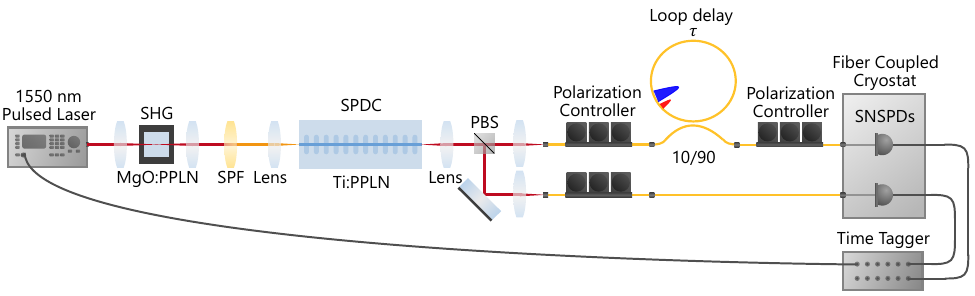}
\caption{Experimental setup for verifying the proposed time filtering approach. Our pump light is generated by second harmonic generation (SHG) pumped by a 1550~nm laser, which is subsequently filtered by a short pass filter (SPF). The resulting 775~nm pump light is then used for generating our single photons in an integrated periodically-poled lithium niobate Type-II SPDC source. The pump light as well as the single TE polarized photons enter the 30~m loop delay through a 10/90 coupler for 1550 nm after passing the polarizing beam splitter (PBS). After exiting the loop system the photons are detected by an SNSPD. The 1550~nm TM polarized idler photons are also detected by an SNSPD in a different arm. These are synchronized with the laser for the overall histogram of the photon pair in the end.}
\label{fig:SetUp}
\end{figure}

To verify our approach, we use a fiber-based loop delay. This allows to show the separation due to the dispersion of the single photons and pump light. We implemented an experiment containing a fiber loop, which allows us to track the separation of the single photons from the pump light at multiple discrete propagation increments. The experimental setup is shown in figure~\ref{fig:SetUp}.

We use periodically poled Ti:LN as an SPDC source which is pumped with 775~nm TE polarized photons and decays into two photons at 1550~nm with TE and TM polarization. The pump light has a power of 10~$\upmu$W which was measured with a photodiode after exiting the source. The laser has a repetition rate of
125 kHz, which leads to a power of 10~pW per pulse.  Afterwards the signal and idler photons are separated by a polarizing beam splitter (PBS). The creation probability  of the source (the fraction of pump pulses in which a down-conversion event occurs) is 61.5~\%. 
The TE polarized signal and pump light are coupled into the fiber loop with a 10/90 coupling ratio as shown in figure~\ref{fig:SetUp}. The high creation probability allows for the creation of multiple photon pairs. Since we are only interested in detectable single photons we still pump at high laser powers in order to reduce the measurement time. The out-coupled light forms a pulse train divided by the time delay~$\mathrm{\tau}$, which is determined by the propagation time of the pulses through the fiber loop. These pulses are detected by an SNSPD. Since the detector is limited by its dead time, a long enough fiber is needed for the loop to ensure the capability of detecting all pulses from the created pulse train. Otherwise the detector would not be able to resolve the pulses and therefore a detection of the single photons would not be possible. The core setup is based on a time-multiplexed detection-experiment~\cite{Tiedau2019}. 

Another challenge lies in the decay of the pulse train, since we are limited by a finite number of time bins to ensure a clear assignment of the pulses. 
The 30~m fiber loop results in a 156.9~ns separation between pulses in the pulse train (also referred to as time bins). The repetition rate of 125 kHz, which leads to an 8~$\upmu$s separation between pump pulses, allows for 51~time bins (i.e., pulses in the pulse train) to be investigated. 
The time tagger records coincidence detection events of an SNSPD signal with the clock signal and compares it afterwards in a combined histogram.

\begin{figure}[t]
\centering\includegraphics{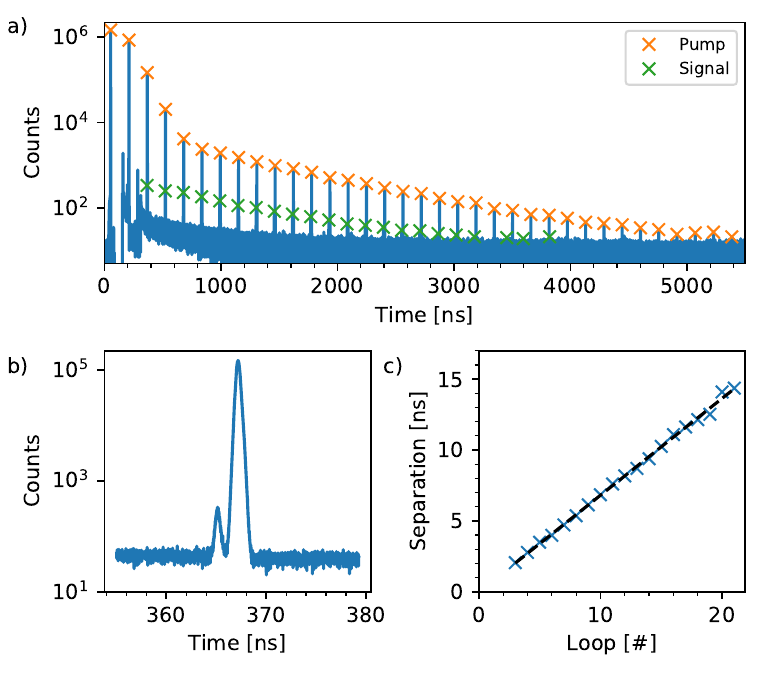}
\caption{(a) Histogram of the clicks of the detected coincidences. The centers of the separated pulses are highlighted by the crosses. The signal (green) and pump (orange) photons separate with each number of round trips spent in the fiber loop while the overall intensity decreases with each round trip. (b) Histogram of the clicks for signal and pump light after three round trips in the fiber loop. The measurement is timed by the laser-clock and heralding photon. (c) Separation in time between the signal photon pulse in contrast to the pump photon pulse for the corresponding amount of round trips.}
\label{fig:Calibright}
\end{figure}

The histogram in figure~\ref{fig:Calibright}~(a) shows the separation of the signal photon pulses relative to the pump pulses over time. The overall intensity of the photon pulses decreases with round trips spent in the delay loop. This results in fewer detected photons and a larger relative detection time between the signal and pump light. Similar measurements with smaller propagation times have already been realized on chip for single wavelengths~\cite{Pernice2012, Ekici2023}.


The first temporal separation of the signal photons from the pump is seen after 3 loops in the delay loop as shown in figure~\ref{fig:Calibright}~(b). This is achieved without any further filtering of the light and still the signal photons and pump light are completely separated. Since the single photons outpace the pump light, we can assume that a click in the single-photon pulse corresponds to a signal photon or dark count of the detector and not to any pump light scattering. This shows a successful separation of pump and signal photons, where the signal is detected clearly ahead of the pump light.

\section{Conclusion}
In conclusion, we have proposed a method to achieve integrated pump suppression for probabilistic photon generation processes, which is based on the photons' relative propagation velocity. Our approach is relevant when the created single photons have a faster propagation velocity than the pump light and therefore arrive first at the detector. In this scenario the detection event would show a single-photon pulse which is completely separated in time from the pump due to its faster propagation speed. The idea of filtering the pulses over time due to their dispersion will be most useful for platforms with single-photon processes where the wavelength of the pump light has a significant dispersion relative to the created single photons and thus a large difference between the effective refractive group indices, which also depends on the implemented platform. This speaks for platforms supporting SPDC such as lithium niobate. However, even platforms with relatively small differences in propagation speed, typical for SFWM, can be useful if propagation losses are sufficiently low, as shown for SiN. 

We emulated this approach by using a 30 m delay fiber loop, which is connected to a 10/90 coupler. Better timing resolution for integrated detectors could reduce the distances until the pulses are separated.

When considering integrated superconducting detectors, these devices are housed in cryogenic environments with limited heat load. One contribution to this is the pump light of the laser, either triggering the detectors, or being absorbed by the cryostat, or appearing as scatter-photons. Our proposed filtering method does provide a separation in the time-domain between single photons and pump photons, however they still are bound to the same waveguides and therefore could be guided out of the cryostat to reduce heating.


\section*{Acknowledgements}

We acknowledge financial support from the Deutsche Forschungsgemeinschaft (Grant No. 231447078–TRR 142) and the Bundesministerium f\"ur Bildung und Forschung (Grant No. 13N14911), as well as the European Union (ERC, Project QuESADILLA, Grant No. 101042399). Views and opinions expressed are however those of the author(s) only and do not necessarily reflect those of the European Union or the European Research Council. Neither the European Union nor the granting authority can be held responsible for them.


\section*{References}
\bibliographystyle{ieeetr}
\bibliography{TimefilteringBib.bib}


\end{document}